\documentclass[preprint,preprintnumbers,amsmath,amssymb]{revtex4}

\usepackage{graphicx}           
\usepackage{dcolumn}            
\usepackage{bm}                 
\usepackage{hyperref}

\begin{document}


\title{Heralded Single-Photon Emission from the Mollow Triplet Sidebands of a Quantum Dot}

\author{A. Ulhaq$^\star$, S. Weiler, S. M. Ulrich, R. Ro{\ss}bach, M. Jetter \& P. Michler}

\affiliation{Institut f\"ur Halbleiteroptik und
Funktionelle Grenzfl\"achen, Universit\"at Stuttgart, Allmandring
3, 70569 Stuttgart, Germany.\\ \\
$^{\star}$~Email correspondence: ata.ulhaq@ihfg.uni-stuttgart.de}

\maketitle


\textbf{Emission from a resonantly excited quantum emitter is a
fascinating research topic within quantum optics and a useful
source for different types of quantum light fields. The resonance
spectrum consists of a single spectral line below saturation of a
quantum emitter which develops into a triplet at powers above
saturation of the emitter
\cite{Mollow:1969,Flagg.Muller:2009,Ates.Ulrich:2009}. The
spectral properties of the triplet strongly depends on pump power
\cite{Ulrich.Ates:2011,Roy.Hughes:2011} and detuning of the
excitation laser. The three closely spaced photon channels from
the resonance fluorescence have different photon statistical
signatures \cite{Schrama.Nienhuis:1992}. We present a detailed
photon-statistics analysis of the resonance fluorescence emission
triplet from a solid state-based artificial atom, i.e. a
semiconductor quantum dot. The photon correlation measurements
demonstrate both 'single' and 'heralded' photon emission from the
Mollow triplet sidebands \cite{Schrama.Nienhuis:1992}. The
ultra-bright and narrowband emission (5.9 MHz into the first lens)
can be conveniently frequency-tuned by laser detuning over 15
times its linewidth ($\Delta\nu \approx 1.0$ GHz). These unique
properties make the Mollow triplet sideband emission a valuable
light source for, e.g.quantum light spectroscopy and quantum
information applications \cite{Kiraz.Atature:2004}.}\\


Generation of non-classical light fields like single-, entangled-
and heralded-photons form a vital part of many schemes of quantum
information and computation. Atom optics demonstrated the heralded
emission of single photons using resonance fluorescence from
single atoms \cite{Aspect.Tannoudji:1980} and from cold atoms in a
cavity \cite{Thompson.Simon:2006}. Currently most of the heralded
photon schemes use sources based on parametric down conversion
\cite{Fasel.Alibart:2004}. Recently, there has been important
developments in fiber-based heralded photon sources using
four-wave mixing \cite{Cohin.Walmsley:2009}. Both of these
techniques present Poissonian statistics with photon bandwidths
usually larger than $100$ GHz. The increased two-photon yield in
these photon sources is at the expense of fidelity. Semiconductor
QDs have demonstrated single- \cite{Michler.Imamoglu:2000},
entangled- \cite{Stevenson.Shields:2006,Akopian.Gershoni:2006} and
heralded-photon sources \cite{Moreau.Gerard:2001} but mostly under
non-resonant excitation schemes. Though, coherent excitation
conditions of those quantum emitters are an important precondition
since these promise to minimize most of the dephasing caused by
the non-resonant, i.e. incoherent processes
\cite{Kiraz.Atature:2004}. Coherent control of QD excitons has
been demonstrated in a variety of experiments exhibiting Rabi
splitting \cite{Kamada:2001}, Rabi oscillations
\cite{Zrenner:2002}, and resonant absorption \cite{Xu.Steel:2007,
Jundt.Imamoglu:2008}. Observations of oscillations in the
first-order correlation \cite{Muller.Flagg:2007}, characteristic
emission spectra in the frequency domain \cite{Vamivakas.Atature:2009},
and oscillations of the second-order photon correlation function
\cite{Flagg.Muller:2009} were all measured directly on the
resonance emission from the QD. In particular, single-photon
emission from a single fluorescence line (below the emitter
saturation) along with photon indistinguishability as high as $90
\%$ was demonstrated \cite{Ates.Ulrich:2009}. Also the AC Stark
shift of an exciton was used to bring the initially split QD
exciton components into resonance with each other, thus forming a
polarization-entangled photon source \cite{Muller.Fang:2009}.
Resonant studies on solid state quantum emitters have also
highlighted some of the fundamental differences vis-a-vis isolated
'real' atoms. It was particularly demonstrated that Mollow spectra
at large Rabi splittings show linewidth broadening due to the
presence of excitation induced dephasing ($EID$) in these quantum
emitters \cite{Ulrich.Ates:2011,{Roy.Hughes:2011}}. Nevertheless,
pure resonant excitation appears to be the ultimate way to achieve
sources of nearly Fourier transform-limited photons preferable to
any non-resonant excitation scheme. Here we extend the
applications of resonance excitation of a QD to demonstrate single
and 'heralded' photon generation from the Mollow triplet
sideband emission.\\

We study the case where resonant pump coherently evolves the
system with 'Rabi' frequency ($\Omega$) larger than the natural
linewidth of the exciton transition ($\Omega \gg \Gamma_{rad}$).
The characteristic resonance fluorescence spectrum of such a
strongly driven system consists of three distinct spectral
components denoted as the 'Mollow triplet' \cite{Mollow:1969}.
These can be identified as the 'Rayleigh' line ($R$) decorated by
spectrally symmetric satellite peaks with the 'fluorescence' line
($F$) as the low energy component and the 'three photon' line
($T$) as the high energy sideband (for terms convention see
\cite{Nienhuis:1993}). These spectral components are a result of
spontaneous emission down a ladder of pairs of so called 'dressed
states'. The dressed states are a representation of the combined
QD state and laser mode along with the coupling between them
\cite{Tannoudji.Grynberg:2004}. Each rung of the ladder (with $N$
number of quanta) consists of two components, i.e. $|1\rangle =
c|g,N+1\rangle + s|e,N\rangle$ and $|2\rangle = c |g,N+1\rangle -
s|e,N\rangle$, where $|g,N+1\rangle$ and $|e,N\rangle$ represent
the eigen-states of the uncoupled quantum emitter and laser
states. The dressed states' components are energetically separated
by the generalized Rabi frequency ($\Omega$) given by the 'bare'
Rabi frequency ($\Omega_0$) and the laser-QD detuning ($\Delta$)
as $\Omega = \sqrt{\Omega_0^2+\Delta^2}$. The amplitudes of the
dressed eigen-states are given as $c = \sqrt{(\Omega +
\Delta)/(2\Omega)}$ and $s = \sqrt{(\Omega - \Delta)/(2\Omega)}$.
The dressed states' eigen-energies as a function of laser detuning
$\Delta$ are plotted in Fig. 1(a).\\


For large positive detunings $|\Delta| \gg 0$, a close inspection
of the steady state solution of these dressed states reveal that
state $|1\rangle$ obtains mainly $|g,N+1\rangle$ character and the
system is mostly prepared in state $|1\rangle$ as $c^2 \gg s^2$
\cite{Schrama.Nienhuis:1992}. Thus a $T$ photon is emitted first
which puts the system into state $|2\rangle$ of the next lower
manifold $E$(N), from where an $F$ photon emission can finally
occur (see Fig. 1(b)). This time-ordered cascaded emission of a
$T$ photon 'heralding' the $F$ transition will result in a photon
bunching signature in photon cross-correlation experiments of the
Mollow sidebands which will be demonstrated below. In contrast, an
autocorrelation measurement on an individual sideband should show
antibunching because after emission of e.g., a $T$ photon,
emission of an $F$ sideband photon should occur before another $T$
photon emission is possible and vice versa (see Figs. 1(b) and
(c)). In contrast, photons from the central Rayleigh line should
be totally uncorrelated in time, i.e. Poissonian distributed, as
the emission of an $R$ photon is not accompanied by population
modulation of the emitter's state \cite{Schrama.Nienhuis:1992}.\\

Figure 1(d) shows a Mollow triplet series obtained at a constant
excitation power under systematic variation of laser detuning
$\Delta$. The central peak shifts along with the laser and the
satellite sidebands always remain spectrally symmetric with
respect to the center. As can be clearly traced in Fig. 1(e), the
total Rabi splitting $\Omega$ increases with increasing detuning
$\Delta$. The linewidth of the Mollow sidebands in the resonant
spectrum ($\Delta = 0$) is found to be $\Delta \nu = 730 \pm
20$~MHz at the corresponding Rabi frequency of $\Omega_0 = 5.4$
GHz.\\


A two steps signal filtering process, initially via a Michelson
interferometer \cite{Aichele.Benson:2004} (see Figs. 2(a)-(c)) and
secondly via a spectrometer, ensures spectral purity of the
photons in Mollow triplet channels. Worth to note, an average
photon count of $70,000$ s$^{-1}$ was obtained on each Mollow
sideband after spectral filtering, which corresponds to a
collection of $5.9$ million photons per second from the sample by
the objective of our confocal microscope.\\

Figure 2(d) shows a photon correlation measurement taken on the
central Mollow peak (see also Fig. 2(c)). Remarkably, the
autocorrelation demonstrates long time scale bunching instead of
pure Poissonian statistics. Such an effect is commonly detected
for QD recombination signal subject to 'blinking' of the excitonic
state between two or more neighboring competing states
\cite{Santori.Pelton:2001}. The timescale of these processes can
vary from $10$ ns to a few hundred ns depending on pump power
\cite{Santori.Pelton:2001}. The phenomenon of blinking under
pure-resonant excitation can be assigned to the presence of a
competing QD exciton spin configuration known as the dark
excitonic state ,which is non-radiative in nature
\cite{Bayer:2002}. Our $g^{(2)}(\tau)$ correlation data is fitted
by a bi-directional exponential fit which reveals the explicit
power-dependent blinking time scale and observed in all the photon
correlation measurements
shown in this work.\\

Results of photon autocorrelation measurement taken on a
spectrally separated Mollow sideband ($T$) are depicted in Fig.
2(e). The whole signal is superimposed by the above-mentioned
bunching-effect due to 'blinking'. The signal around zero delay
exhibits clear antibunching with a normalized value of
$g^{(2)}(\tau) = 0.18$ at $\tau = 0$, considering the time
resolution of the detection setup. From the experimental signature
of antibunching, we derive a time constant of $\tau_{fit} = 0.8
\pm 0.1$ ns, which is somewhat smaller than the theoretically
expected value \cite{Schrama.Nienhuis:1992} (see supplemental
material). We interpret the presence of a finite background in the
correlation measurement as a consequence of signal contribution
from the spectrally-close central peak of the Mollow
triplet as non-vanishing background in the PL spectrum (see inset Fig. 2(e)).\\


Figure 3(a) demonstrates excerpts of a series of correlation
measurements taken on both the sidebands simultaneously without
spectral separation between them. Consequently, there is no time
ordering between the sidebands photon emission as either of them
can initiate a 'start' or 'stop' signal in the photon counting
process. The data of Fig. 3(a) represents three distinct cases,
i.e. the laser near the exact QD s-shell ($\Delta \approx 0$;
center), blue-detuned ($\Delta > 0$; top) and red-detuned ($\Delta
< 0$; bottom), from the resonance. As expected a short time scale
\textit{symmetric} bunching feature is clearly observed on both
the detuned cases $\Delta \neq 0$. The resonant case $\Delta
\approx 0$ doesn't show such bunching as the probability of both
the $F$ and $T$ as a first photon in the cascade is equal and
hence no heralded emission is expected. Again, all the three
measurements reveal the long time scale ($\Gamma_{bl}^{-1} = 42$
ns here) bunching effect corresponding to the 'blinking' of the
state. Figure~3(b) displays the measured evolution of the bunching
strength as a function of the detuning $\Delta$ for two different
fixed bare Rabi energies $\Omega_0$. A symmetric increase of the
bunching signature to either side of the detuning is observed as
is obvious from the dashed lines as guide to the eye in the
figure. Even though the 'heralding' nature of photon emission is
observed, the time ordering is still not observed
due to the simultaneous detection of both sidebands.\\


Figure 4(a) and (b) show cross-correlation measurements between
the spectrally separated opposite sidebands of the Mollow triplet
in which the $T$ photon always act as a 'start' and the $F$ photon
as a 'stop' signal, respectively. In each case, a short time scale
bunching with a time \textit{asymmetric} peak is now observed.
Fig. 4(a) shows the case of a strongly blue-detuned
($|\Delta|/\Omega \sim 83\%$) laser with an abrupt rising of the
bunching signal ($0.48$ ns, equivalent to the response time of
detectors) for negative delay times and a longer exponential decay
(0.94 ns) for positive delays. This directly reflects that each
$F$  photon is 'heralded' by a $T$ photon (see Fig. 1(b)). Fig.
4(b) shows the case of a strongly red-detuned ($|\Delta|/\Omega
\sim 79\%$) laser where the time ordering is now reversed, i.e.
here the $T$ photon is heralded by the $F$ photon (see Fig. 1(c)).
The falling edge of the correlation functions can be perfectly
fitted by the time scales ($0.94 \pm 0.10$ ns for blue-detuned;
$0.89 \pm 0.10$~ns for red-detuned case) close to the
theoretically predicted photon emission rates given by $\Gamma =
\Gamma_{rad} (c^4+s^4)$ \cite{Schrama.Nienhuis:1992}, yielding
values of $1.10 \pm 0.13$ ns  (blue-detuned) and $1.15 \pm
0.10$~ns
(red-detuned), respectively.\\

In conclusion, we have demonstrated that an individual Mollow
sideband channel of the resonance fluorescence from a single
quantum dot can act as an efficient single-photon source. By
spectrally separating both the Mollow sidebands, we showed that
this source can act as a solid state-based heralded photon emitter
as well. Possible applications of such high-brightness photon
sources include the realization of semiconductor-atom interfaces
for the development of photon based-memories
\cite{Akopian.Rastelli:2011}. Heralded photon sources may find
applications in quantum communication protocols
\cite{Gisin.Thew:2007}. These high yield and narrow band sources
might also be invaluable for spectroscopy of quantum emitters,
where non-classical light is used for the excitation process.
Using the Purcell effect by embedding the QDs in optical
microcavities can result in spectrally ultrasharp emission lines
\cite{Freedhoff.Quang:1994}. Tuning the cavity into resonance with
one of the Mollow triplet sideband can result in population
inversion of the system and ultimately single quantum emitter
lasing
\cite{Quang. Freedhoff:1993}.\\

\subsection*{Method Summary}
The sample consist of a single layer of self-assembled In(Ga)As
quantum dots grown by metal-organic vapor-phase epitaxy (MOVPE).
The quantum dot layer is sandwiched between GaAs/AlAs DBR layers
at the center of a GaAs $\lambda$-cavity. The sample is kept in a
Helium-flow cryostat which is capable to stabilize the temperature
to $5 \pm 0.5$ K. The measurements are performed on a special
micro-Photoluminescence ($\mu$PL) setup \cite{Ates.Ulrich:2009}
with orthogonal geometry between the laser excitation in the
growth plane, in combination with detection of luminescence
perpendicular to the sample surface. The DBR structure of the
sample acts as a waveguide for the excitation laser. Additional to
$90^{\circ}$ geometry, a pinhole assembly is employed to minimize
the contribution of scattered laser light. A polarizer setup
containing a high-extinction Glan Thompson polarizer further
suppresses the collection of laser stray light via polarization
selection. An individual QD is addressed by a narrow-band ($\sim
500$ kHz) cw Ti:Sa laser by tuning it into the s-shell of a QD.
While scanning the narrowband laser over the QD s-shell resonance,
enhancement of the signal is an indication of the onset of
resonance fluorescence. The spectrometer + CCD combination in the
setup has a resolution of 8.4~GHz, which cannot resolve the
components of Mollow emission spectrum particularly at lower
excitation powers. To obtain a detailed spectrum of the full
Mollow triplet, with well-resolved sidebands, we use a scanning
Fabry-P\'{e}rot interferometer
providing a resolution of less than $\sim 250$ MHz.\\

In order to perform correlation measurements on different spectral
components of the Mollow triplet, it is important to have an
efficient filtering mechanism. For this purpose, a Michelson
interferometer is employed for pre-filtering. The delay of the
interferometer is kept in such a way that the photons from Mollow
sidebands interfere constructively on one output, while the
central Mollow peak leave the other output of the interferometer.
After the spatial separation of the sidebands and the central
peak, further filtering is performed via two spectrometers with
spectral resolutions of $8.4$ and $9.6$ GHz, respectively. The
same spectral component of the Mollow emission spectrum is
selected to perform photon auto-correlation, while for
cross-correlations different components are selected by each spectrometer.\\


\subsection*{Acknowledgements}

The authors greatly acknowledge D. Richter and W.-M. Schulz for
providing high quality sample, and M. Wiesner for help in sample
processing. The authors gratefully acknowledge financial support
by the DFG research group 730. A. Ulhaq acknowledges funding from
International Max Planck Research School IMPRS-AM. S. Weiler
acknowledges financial support by the Carl-Zeiss-Stiftung.

\subsection*{Author contributions}

R.R. and M.J. designed the sample structure. A.U., S.W., S.M.U.
and P.M. conceived the experiments. A.U., S.W. and S.M.U.
performed the experiments and analyzed the data. A.U., S.W.,
S.M.U. and P.M. wrote the article, with input from the other
co-authors.

\newpage


\textbf{Figure 1: Laser-detuning dependent Mollow triplet
spectra.} (a) Laser-QD-detuning dependence of dressed state
eigen-energies (solid lines), together with the QD + laser
uncoupled eigenstates (dashed lines)
\cite{Tannoudji.Grynberg:2004}. (b) and (c) Sideband photon
emission sequence down the dressed states ladder for a strong
blue-detuned ($\Delta \gg 0$,(b)) and a red-detuned ($\Delta \ll
0,$(c)) laser case. (d) Laser detuning-dependent resonance
fluorescence spectra recorded in high-resolution PL. All the
spectra are taken at the same excitation power of $P = 500$
$\mu$W. (e) Positions of the spectral components of the
fluorescence signal derived from the detuning series in (d). Solid
lines are fits according to $\nu = \Delta \pm \sqrt{\Omega_0^2 +\Delta^2}$ and $\nu = \Delta$. \\

\textbf{Figure 2: Photon correlations on Mollow spectral
components.} (a) Michelson interferometer setup used to spatially
separate the Mollow sidebands from the central line (see
supplementary information). (b) Mollow spectrum obtained after
filtering out the central peak, where only the sidebands remain
observable in high-resolution PL. (c) High-resolution spectrum of
the central Rayleigh line filtered from the Mollow sidebands in
(b). (d) Photon autocorrelation of the central Mollow triplet
peak. The 'bare' Rabi frequency is $11.27$ GHz while the laser is
at exact resonance with the QD exciton ($\Delta = 0$).  The solid
line is a bi-directional exponential fit to the data to extract
the effect of 'blinking'. Inset: PL spectrum of the full Mollow
triplet obtained with the shaded region as the selected spectral
region for photon correlation measurement. (e) Auto-correlation on
the $T$ line of the Mollow spectrum. The experimental conditions
were $\Omega_0 = 11.27$~GHz and $\Delta = 0$. The solid red (blue
dashed) line is a theoretical fit of the signal, convoluted
(deconvoluted) with the instrumental detector response ($480$ ps).
Inset: PL spectrum after filtering out the central line. The
shaded region shows the spectrally selected
signal for the autocorrelation measurement. \\

\newpage

\textbf{Figure 3: Laser detuning-dependent correlation of the
combined Mollow sidebands signal.} (a) Photon correlation
measurement on both Mollow triplet sidebands, without any
time-ordering. The measurements are performed on resonance with
the QD s-shell (center), blue detuned (top) and red-detuned
(bottom). Inset: Corresponding PL spectrum with highlighted area
as the spectrally selected signal. (b) Bunching parameter, derived
from correlation measurements (a), as a function of laser-QD
detuning $\Delta$. The data points are plotted for two different
excitation powers: red ($170$ $\mu$W with $\Omega_0 = 3.4$ GHz)
and black ($500 \mu$W with $\Omega_0 = 5.4$ GHz).
Dashed lines are guides to the eye.\\

\textbf{Figure 4: Cross-correlation between spectrally selected
Mollow sidebands.} (a) Photon correlation between the $T$ photon
('start') and $F$ photon ('stop') lines, where the central
Rayleigh line is interferometrically suppressed. The 'bare' Rabi
frequency is $\Omega_0 = 11.1$ GHz, while the laser is
blue-detuned by $\Delta = +9.3$ GHz from exact resonance. The
solid line is a theoretical fit to the data with the exponential
for negative (positive) delay indicating the rising (falling) edge
of the correlation signal. For this case, the steady state
population of $|1\rangle$ is $95 \%$. (b) Same as (a) but for a
case where $\Omega_0 = 9.6$ GHz while $\Delta = -7.6$ GHz. The
rise and falling edge are time-reversed in this case. The steady
state population of state $|2\rangle$ is $95 \%$ in this case.

\newpage


\textbf{Figure~1}

\begin{figure}[!ht]
\begin{centering}
\includegraphics[width=13cm]{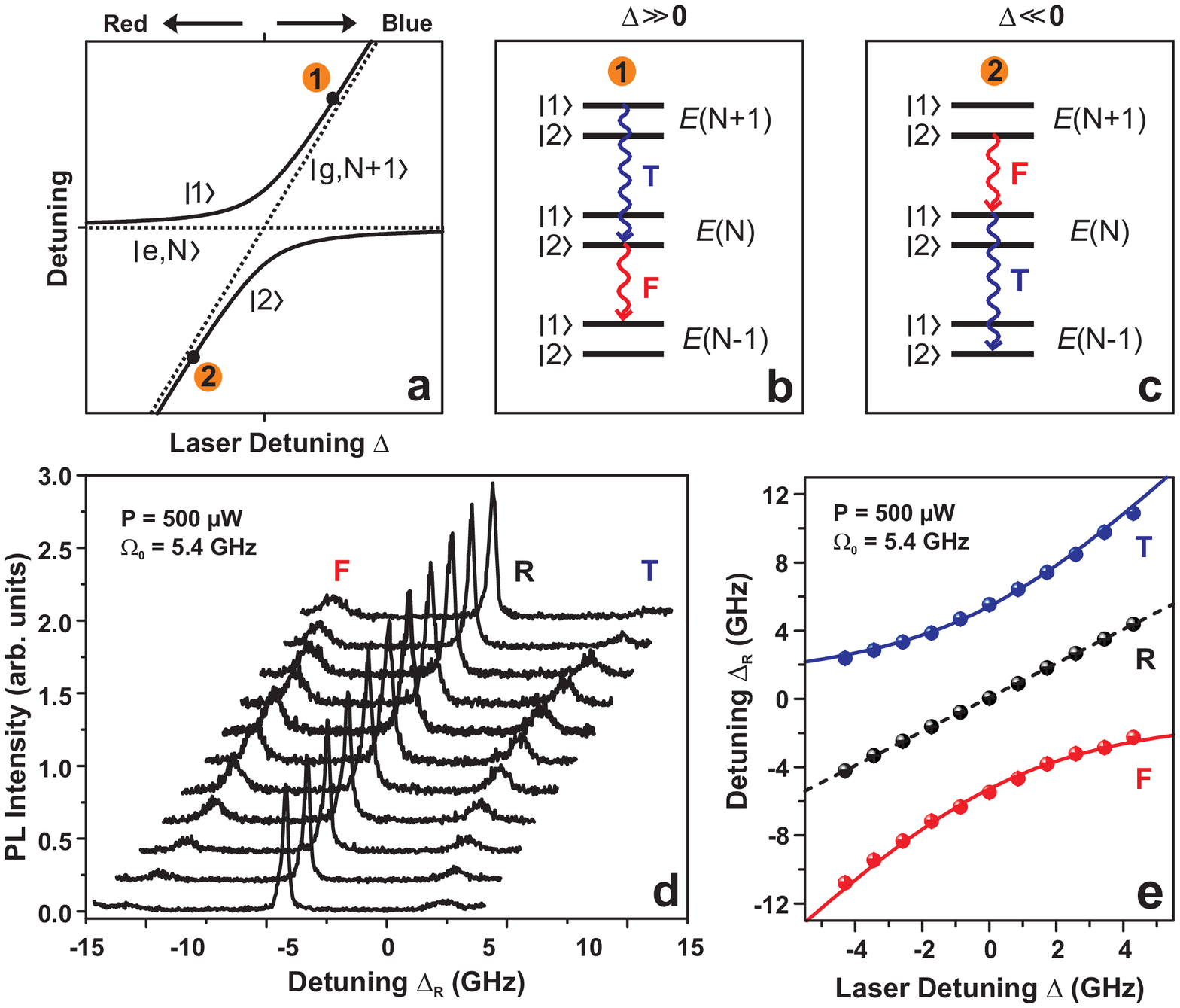}
\label{fig:1}
\end{centering}
\end{figure}

\newpage
\textbf{Figure~2}

\begin{figure}[!ht]
\begin{centering}
\includegraphics[width=13cm]{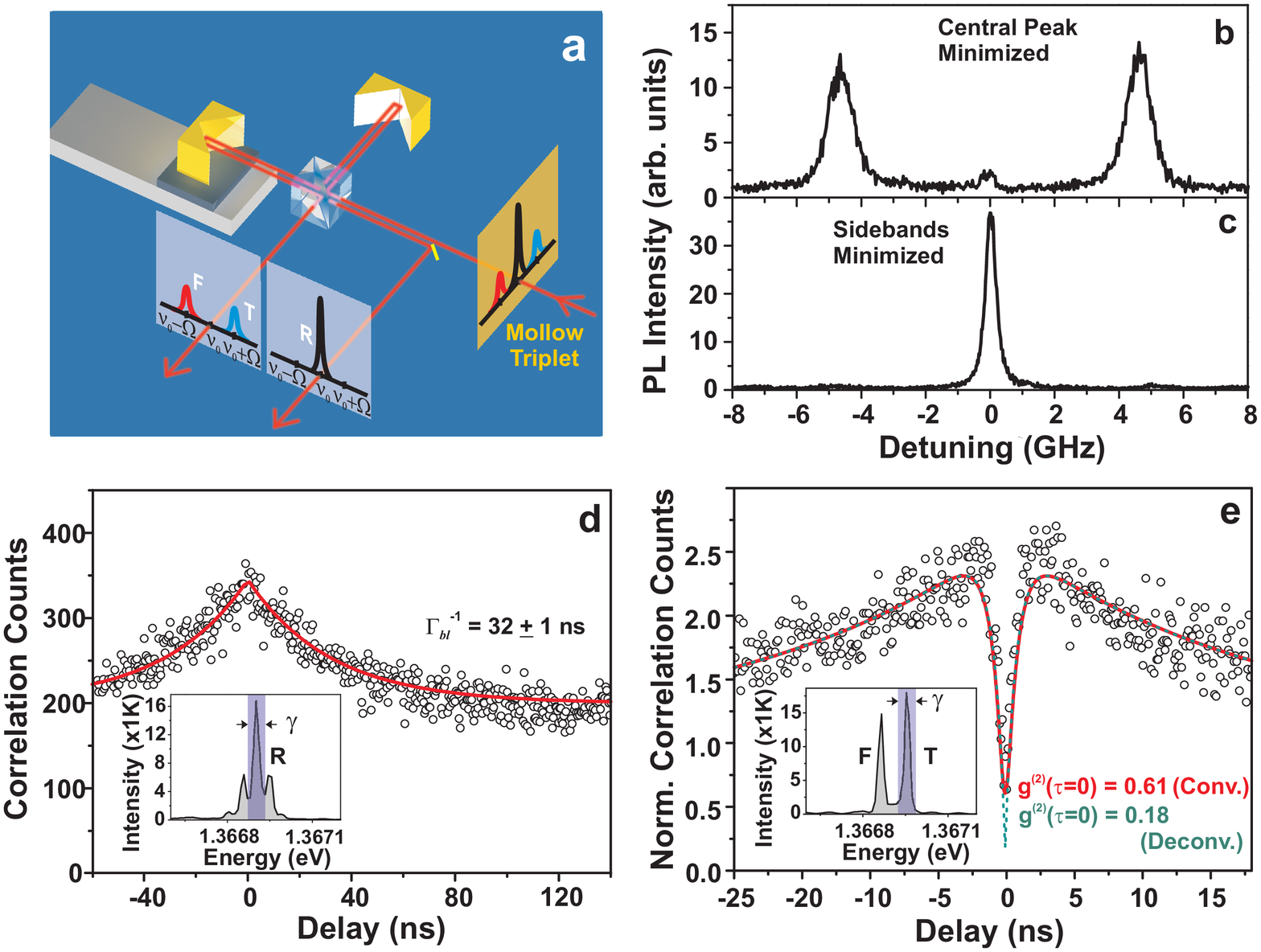}
\label{fig:2}
\end{centering}
\end{figure}

\newpage
\textbf{Figure~3}

\begin{figure}[!ht]
\begin{centering}
\includegraphics[width=15cm]{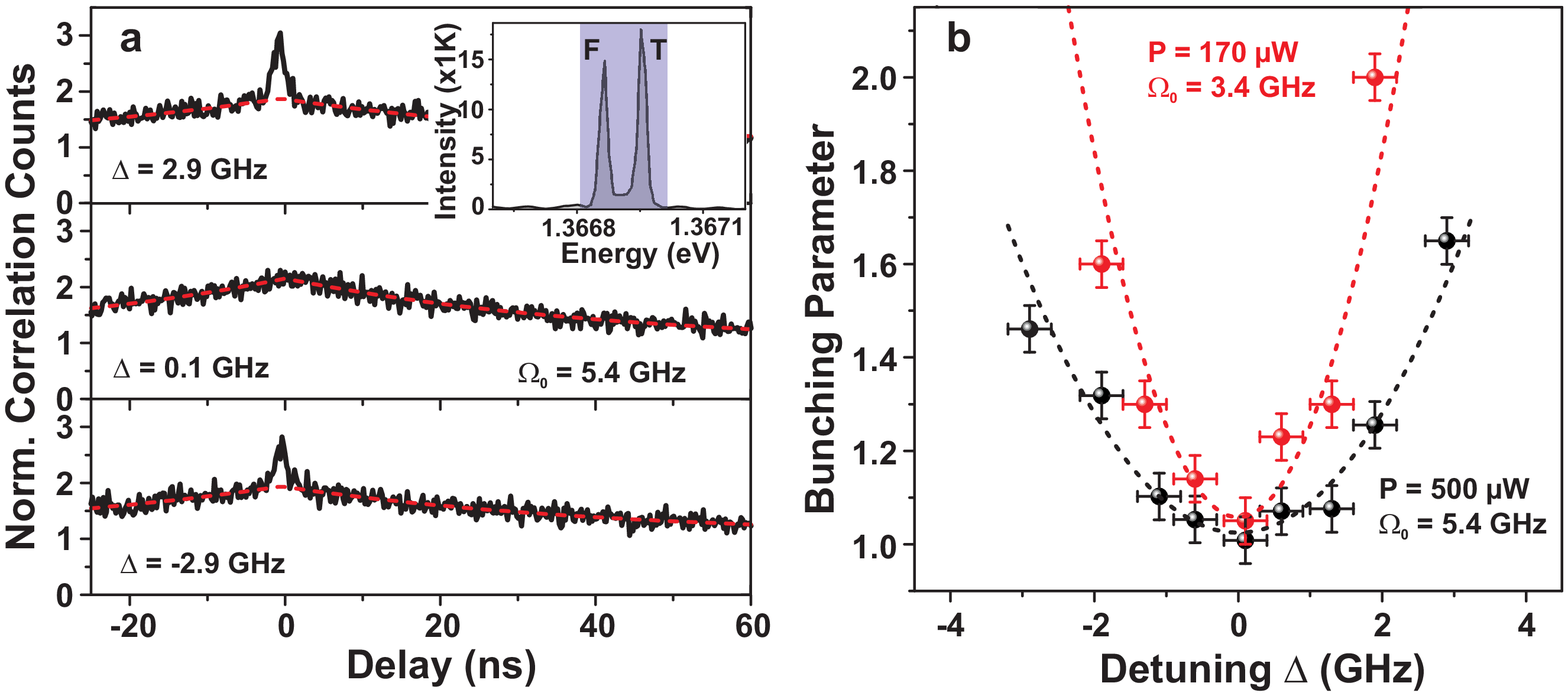}
\label{fig:3}
\end{centering}
\end{figure}

\newpage
\textbf{Figure~4}

\begin{figure}[!ht]
\begin{centering}
\includegraphics[width=15cm]{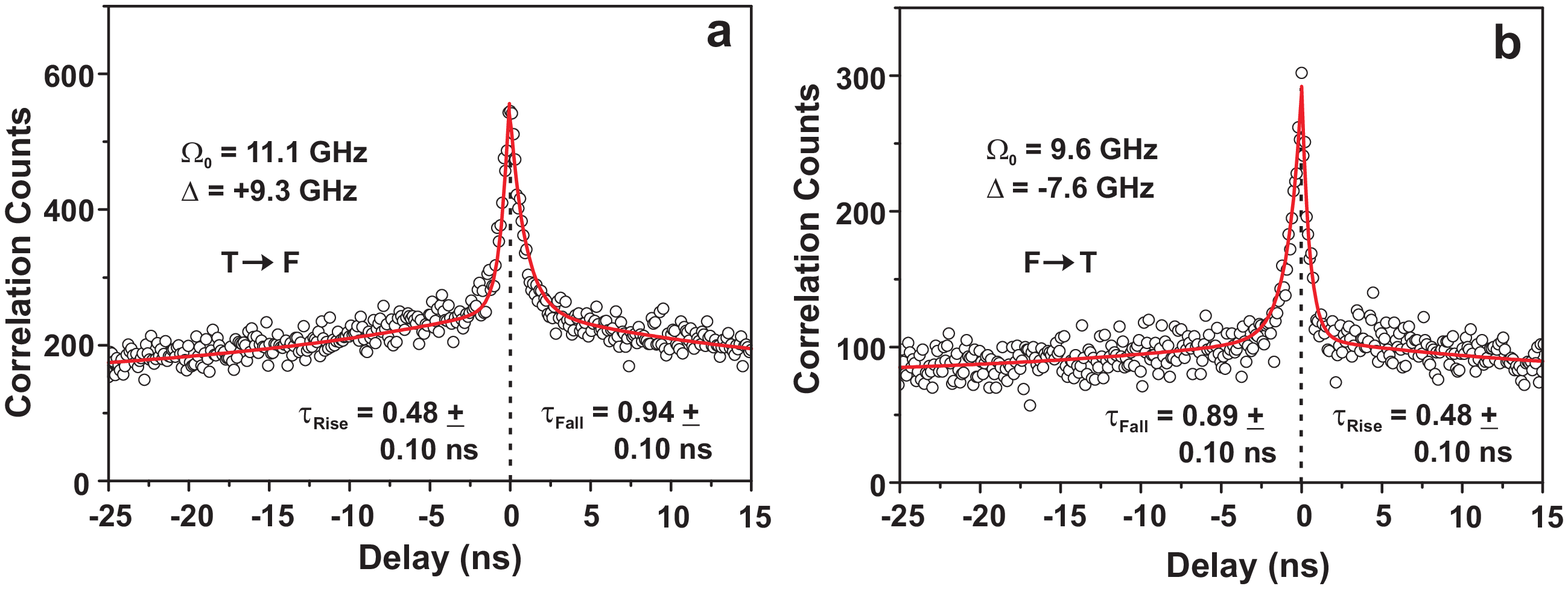}
\label{fig:4}
\end{centering}
\end{figure}

\newpage

\section*{Supplementary Data}
In the supplementary data we provide further explanation of
experimental methods and conditions. Some of the analysis has been
explained and additional results are added, which give deeper
insight of the concepts in the main text.

\subsection*{1. Methods}


The Michelson interferometer shown in Fig. 2(a) is used to
selectively suppress the sidebands or the central line of the
Mollow triplet. As explained in the main text for the measurements
presented in Figs. 2 (d) and (e), the sidebands or the central
peak are suppressed, respectively . The probability of photons of
a certain wavelength $\lambda$ to leave one output of the
interferometer beam splitter is given by \cite{Aichele:2004}:
\begin{equation}\label{eqn:ModTime}
p(\lambda) = 0.5 + 0.5*cos(2\pi d/\lambda)
\end{equation}
where $d$ is the path length difference between the two arms of
the interferometer. The path difference should be well below the
coherence length of the emission. The minimum path difference for
which the two components with spectral separation of $\Delta
\lambda$ are spatially separated is given by $d_0 = \lambda
(\lambda + \Delta\lambda)/(2\Delta\lambda)$. Figure S1 plots the
photon output probability of all the three components of the
Mollow triplet for a Rabi splitting of $10$ GHz ($\Delta \lambda =
0.027$ nm) around $d_0 = 15.4947$ mm (vertical black dashed line).
The sideband photons (red solid and blue dashed lines) interfere
constructively at $d_0$, while the central peak photons (black
line) interferes destructively at the same output of the beam
splitter. It can be readily observed that at the specific path
difference (vertical dashed line) the output beam consists mainly
of the sidebands and almost negligible contribution from the
central peak. In contrast, the other output of the interferometer
will mainly consists of the central Mollow peak. A motorized stage
with a linear step resolution of $100$ nm is used to accurately
adjust the output of the Michelson interferometer.\\

For the demonstration of cross-correlation between the two
sidebands, it is important that the time ordering of the heralded
photon pairs is maintained by the photon detection process as
well. A spectral filter of bandwidth $\gamma$ introduces a time
uncertainty of $1/\gamma$ in the filtered photons. Thus the time
ordering of the detection of each spectral component will be
maintained only if the photon arrival-time-difference $\tau$ is
larger than the inverse bandwidth ($\gamma^{-1}$) of the filter
($110$ ps $\leftrightarrow$ $35$ $\mu$eV)
\cite{Nienhuis.Nienhuis:1993}. The timescale between the sideband
photons is given by the modified lifetime ($\Gamma^{-1}$) which is
greater $\gamma^{-1}$ for the experimental data shown in Fig. 4(d)
and (e) of the main text. It's noteworthy that this formulation is
valid
as long as the inequality $\Omega \gg \gamma \gg \Gamma_{rad}$ is satisfied.\\

\subsection*{2. Data Analysis}

There are two kinds of bunching effects in Figures 3 and 4, where
one is a long time scale effect due to the blinking of the QD dark
state as explained in the main text. These are normally longer
than $10$ ns. The other bunching is due to the heralded photon
emission and appears on a much shorter timescale of the modified
emission lifetime ($\Gamma^{-1}$) of the QD exciton. The
theoretical fit to these data plots are given by:
\begin{equation}\label{eqn:ModTime}
g^{(2)}(\tau) = 1 + c_1^2 e^{-\Gamma\tau} + c_2^2 e^{-\Gamma_{bl}\tau}
\end{equation}
The timescale of the blinking effect has a clear pump power
dependence such that its time scale decreases with increasing
excitation power \cite{Santori.Salamo:2001}.\\
As mentioned in the main text, the modified emission rate $\Gamma$
for the heralded emission from the Mollow sidebands under explicit
laser-QD detuning $\Delta$ is calculated by the relation:
\begin{equation}\label{eqn:ModTime}
\Gamma = \Gamma_{rad} (c^4+s^4).
\end{equation}
For the auto-correlation measurement in Fig.~2(e), the following
parameters are expected for the conditions of zero detuning
$\Delta = 0$: $c^2 = 0.5$, $s^2 = 0.5$, and $\Gamma = 1.6 \pm
0.1$~GHz. \\ For the experimental conditions in Figs. 4(a) and (b), the following parameters are derived:\\
Blue-detuned case (Fig. 4(a)): $c^2 = 0.82 \pm 0.04$; $s^2 = 0.18 \pm 0.04$; $\Gamma_{rad} = 1.25 \pm 0.08$ GHz, $\Gamma = 1.13 \pm 0.10$~GHz.\\
Red-detuned case (Fig. 4(b)): $c^2 = 0.19 \pm 0.04$; $s^2 = 0.81 \pm 0.04$; $\Gamma_{rad} = 1.25 \pm 0.08$ GHz, $\Gamma = 1.15 \pm 0.10$~GHz.\\

\subsection*{3. Additional measurements (p-shell excitation)}

For these measurements the QD was selectively pumped via an
excited state (p-shell). The PL spectrum under these conditions
show a clear exciton line from the QD (see inset of Fig. S2(a)).
Figure S2(a) depicts the time correlated single photon counting
(TCSPC) data from the QD. The measurement was performed through a
fast response APD (response time of $\sim 45$ ps). The data is
fitted to a single exponential and yields a radiative lifetime of
$800 \pm 10$ ps of the QD exciton. Figure S2 (b) presents an
auto-correlation measurement on the QD emission signal. The
correlation measurement under p-shell excitation also exhibits the
long time scale ($\sim 49$ ns) 'blinking' effect resulting in a
bunching of the overall data. The data points near zero delay
demonstrate clear antibunching on a fast time scale (on the order
of radiative emission lifetime), clearly revealing the single
photon nature of the QD line. The deconvoluted fit to the data
points yields an antibunching value of $0.23$. The finite
background can be due to a spectrally close line from another QD
(see inset of Fig. S2(a)). The data has been fitted using the
following theoretical function:
\begin{equation}\label{eqn:ModTime}
g^{(2)}(\tau) = 1 - c_1^2 e^{-\Gamma_{rad}\tau} + c_2^2 e^{-\Gamma_{bl}\tau}
\end{equation}
where $\Gamma_{rad}^{-1}$ is the spontaneous emission lifetime and
$\Gamma_{bl}^{-1}$ is the extracted timescale associated with the
'blinking' effect of the QD. The coefficients $c_1$ and $c_2$
represents the signal to noise ratio of the detected signal. The
same kind of fit is used for data in Fig 2(e) of the main text.


\newpage


\textbf{Figure S1: Michelson interferometer-based frequency filtering.}
The probability of emission of photons at an output of the interferometer
as a function of varying path difference between the two
arms of the interferometer. The black solid line is the probability of the Mollow
central line while the red line is the probability of the red-sideband and
blue dashed line is the probability of blue sideband of the Mollow triplet.
The blue line has been shifted vertically for better observation. The central line
is taken at $907$ nm ($330531.9$ GHz) while a Rabi splitting of $10$ GHz is assumed. \\ \\

\textbf{Figure S2: Time-resolved and photon correlation measurement under quasi-resonant excitation. }
(a) Inset shows the spectrum of the QD under p-shell excitation. The data shows a
time correlated photon counts from the QD line under pulsed p-shell excitation. A single
exponential fit extracts the emission lifetime of the QD. (b) Autcorrelation measurement on the
QD signal under quasi-resonant excitation. The solid red (dashed blue) line is a theoretical
fit to the data convoluted (deconvoluted) with the instrument response function.\\

\newpage


\textbf{Figure~S1}

\begin{figure}[!ht]
\begin{centering}
\includegraphics[width=9cm]{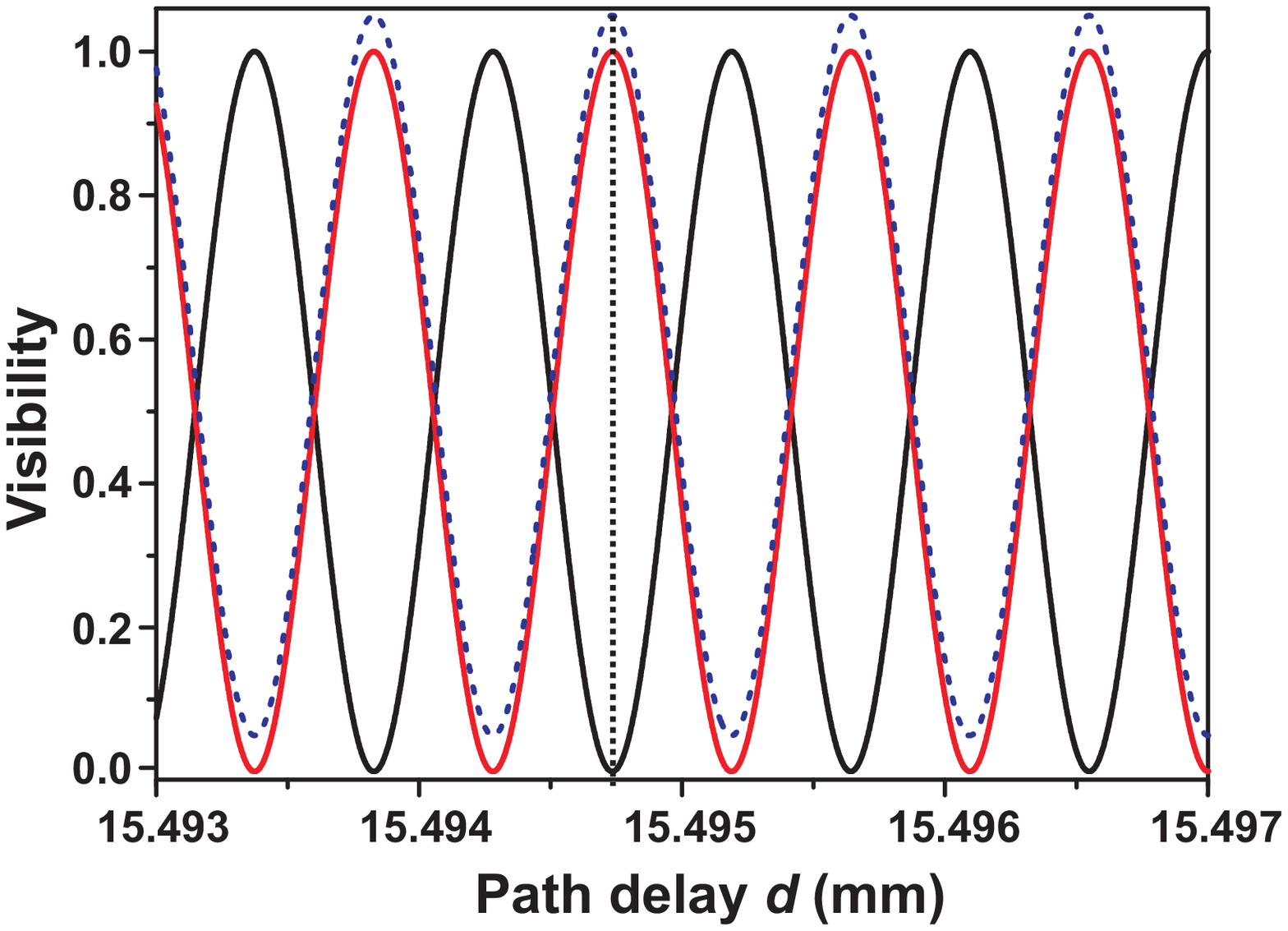}
\label{fig:1}
\end{centering}
\end{figure}

\newpage
\textbf{Figure~S2}

\begin{figure}[!ht]
\begin{centering}
\includegraphics[width=15cm]{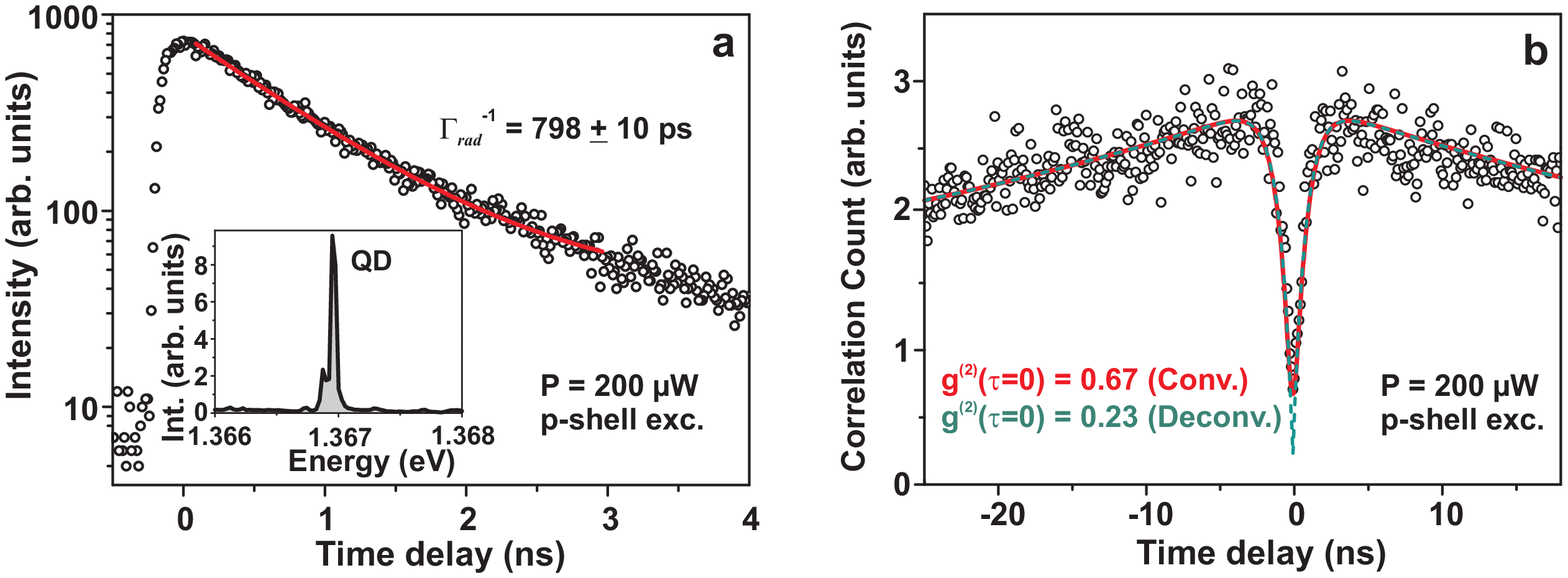}
\label{fig:2}
\end{centering}
\end{figure}


\begin{thebibliography}{10}
\expandafter\ifx\csname url\endcsname\relax
  \def\url#1{\texttt{#1}}\fi
\expandafter\ifx\csname
urlprefix\endcsname\relax\def\urlprefix{URL}\fi
\providecommand{\bibinfo}[2]{#2}
\providecommand{\eprint}[2][]{\url{#2}}

\bibitem{Mollow:1969}
\bibinfo{author}{Mollow, B. R.}
\newblock \bibinfo{title}{Power Spectrum of Light Scattered by Two-Level Systems}.
\newblock \emph{\bibinfo{journal}{Phys. Rev.}}
  \textbf{\bibinfo{volume}{188}}, \bibinfo{pages}{1969}
  (\bibinfo{year}{1969}).

\bibitem{Flagg.Muller:2009}
\bibinfo{author}{Flagg, E. B.}, \emph{et al.}
\newblock \bibinfo{title}{Resonantly driven coherent oscillations in a solid-state quantum emitter}.
\newblock \emph{\bibinfo{journal}{Nature Physics}}
  \textbf{\bibinfo{volume}{5}}, \bibinfo{pages}{203-207}
  (\bibinfo{year}{2009}).

\bibitem{Ates.Ulrich:2009}
\bibinfo{author}{Ates, S.}, \emph{et al.}
\newblock \bibinfo{title}{Post-Selected Indistinguishable Photons from the Resonance Fluorescence of a Single Quantum Dot in a Microcavity}.
\newblock \emph{\bibinfo{journal}{Phys. Rev. Lett.}}
  \textbf{\bibinfo{volume}{103}}, \bibinfo{pages}{167402}
  (\bibinfo{year}{2009}).

\bibitem{Ulrich.Ates:2011}
 \bibinfo{author}{Ulrich, S. M.}, \emph{et al.}
\newblock \bibinfo{title}{Dephasing of Mollow Triplet Sideband Emission of a Resonantly Driven Quantum Dot in a Microcavity}.
\newblock \emph{\bibinfo{journal}{Phys. Rev. Lett.}}
 \textbf{\bibinfo{volume}{106}}, \bibinfo{pages}{247403}
  (\bibinfo{year}{2011}).

\bibitem{Roy.Hughes:2011}
\bibinfo{author}{Roy, C.} \& \bibinfo{author}{Hughes, S.}
\newblock \bibinfo{title}{Phonon-dressed Mollow triplet in the regime of cavity-QED}.
\newblock \emph{\bibinfo{journal}{Phys. Rev. Lett.}}
  \textbf{\bibinfo{volume}{106}}, \bibinfo{pages}{247402}
  (\bibinfo{year}{2011}).

\bibitem{Schrama.Nienhuis:1992}
\bibinfo{author}{Schrama, A.}, \bibinfo{author}{Nienhuis, G.},
  \bibinfo{author}{Dijkerman, H. A.}, \bibinfo{author}{Steijsiger, C.}
   \& \bibinfo{author}{Heideman, H. G.}
\newblock \bibinfo{title}{Intensity correlations between the components of the resonance fluorescence triplet}.
\newblock \emph{\bibinfo{journal}{Phys. Rev. A}}
  \textbf{\bibinfo{volume}{45}}, \bibinfo{pages}{8045}
  (\bibinfo{year}{1992}).

\bibitem{Kiraz.Atature:2004}
\bibinfo{author}{Kiraz, A.}, \bibinfo{author}{Atat\"ure, M.} \&
  \bibinfo{author}{Imamo\u{g}lu}.
\newblock \bibinfo{title}{Quantum-dot single-photon sources: Prospects for
  applications in linear optics quantum-information processing}.
\newblock \emph{\bibinfo{journal}{Phys. Rev. A}} \textbf{\bibinfo{volume}{69}},
  \bibinfo{pages}{032305} (\bibinfo{year}{2004}).

\bibitem{Aspect.Tannoudji:1980}
\bibinfo{author}{Aspect, A.}, \bibinfo{author}{Roger, G.},
  \bibinfo{author}{Reynaud, S.}, \bibinfo{author}{Dalibard, J.}
   \& \bibinfo{author}{Cohen-Tannoudji, C.}
\newblock \bibinfo{title}{Time Correlations between the Two Sidebands of the Resonance Fluorescence Triplet}.
\newblock \emph{\bibinfo{journal}{Phys. Rev. Lett.}}
  \textbf{\bibinfo{volume}{45}}, \bibinfo{pages}{617-620}
  (\bibinfo{year}{1980}).

\bibitem{Thompson.Simon:2006}
\bibinfo{author}{Thompson, J. K.}, \bibinfo{author}{Simon, J.},
\bibinfo{author}{Huanquian, L.} \& \bibinfo{author}{Vuleti\'{c}, V.}
\newblock \bibinfo{title}{A High-Brightness Source of Narrowband, Identical-Photon Pairs}.
\newblock \emph{\bibinfo{journal}{Science}}
  \textbf{\bibinfo{volume}{313}}, \bibinfo{pages}{74-77}
  (\bibinfo{year}{2006}).

\bibitem{Fasel.Alibart:2004}
\bibinfo{author}{Fasel, S.}, \emph{et al.}
\newblock \bibinfo{title}{High-quality single asynchronous heralded single-photon source at telecom wavelength}.
\newblock \emph{\bibinfo{journal}{N. J. Phys.}}
  \textbf{\bibinfo{volume}{6}}, \bibinfo{pages}{163}
  (\bibinfo{year}{2004}).

\bibitem{Cohin.Walmsley:2009}
\bibinfo{author}{Cohin, O.}, \emph{et al.}
 \newblock \bibinfo{title}{Tailored Photon-Pair Generation in Optical Fibers}.
\newblock \emph{\bibinfo{journal}{Phys. Rev. Lett.}}
  \textbf{\bibinfo{volume}{102}}, \bibinfo{pages}{123603}
  (\bibinfo{year}{2009}).

\bibitem{Michler.Imamoglu:2000}
\bibinfo{author}{Michler, P.}, \emph{et al.}
\newblock \bibinfo{title}{A Quantum Dot Single-Photon Turnstile Device}.
\newblock \emph{\bibinfo{journal}{Science}}
  \textbf{\bibinfo{volume}{290}}, \bibinfo{pages}{2282-2285}
  (\bibinfo{year}{2000}).

\bibitem{Akopian.Gershoni:2006}
\bibinfo{author}{Akopian, N.}, \emph{et al.}
\newblock \bibinfo{title}{Entangled Photon Pairs from Semiconductor Quantum Dots}.
\newblock \emph{\bibinfo{journal}{Phy. Rev. Lett.}}
  \textbf{\bibinfo{volume}{96}}, \bibinfo{pages}{130501}
  (\bibinfo{year}{2006}).

\bibitem{Stevenson.Shields:2006}
\bibinfo{author}{Stevenson, R. M.}, \emph{et al.}
\newblock \bibinfo{title}{A semiconductor source of triggered entangled photon pairs}.
\newblock \emph{\bibinfo{journal}{Nature}}
  \textbf{\bibinfo{volume}{439}}, \bibinfo{pages}{179-182}
  (\bibinfo{year}{2006}).

\bibitem{Moreau.Gerard:2001}
\bibinfo{author}{Moreau, E.}, \emph{et al.}
\newblock \bibinfo{title}{Quantum Cascade of Photons in Semiconductor Quantum Dots}.
\newblock \emph{\bibinfo{journal}{Phys. Rev. Lett.}}
  \textbf{\bibinfo{volume}{87}}, \bibinfo{pages}{183601}
  (\bibinfo{year}{2001}).

\bibitem{Kamada:2001}
\bibinfo{author}{Kamada H}, \emph{et al.}
\newblock \bibinfo{title}{Exciton Rabi oscillation in a single quantum dot}.
\newblock \emph{\bibinfo{journal}{Phys. Rev. Lett.}}
  \textbf{\bibinfo{volume}{87}}, \bibinfo{pages}{246401}
  (\bibinfo{year}{2001}).

\bibitem{Zrenner:2002}
\bibinfo{author}{Zrenner, A.}, \emph{et al.}
\newblock \bibinfo{title}{Coherent properties of a two-level system based on a quantum-dot photodiode}.
\newblock \emph{\bibinfo{journal}{Nature}}
  \textbf{\bibinfo{volume}{418}}, \bibinfo{pages}{612-614}
  (\bibinfo{year}{2002}).

\bibitem{Xu.Steel:2007}
\bibinfo{author}{Xu, X.}, \emph{et~al.}
 \newblock \bibinfo{title}{Coherent Optical Spectroscopy of a Strongly Driven Quantum Dot}.
\newblock \emph{\bibinfo{journal}{Science}}
  \textbf{\bibinfo{volume}{317}}, \bibinfo{pages}{929-932}
  (\bibinfo{year}{2007}).

\bibitem{Jundt.Imamoglu:2008}
\bibinfo{author}{Jundt, G.}, \emph{et al.}
 \bibinfo{author}{Robledo, L.},
  \bibinfo{author}{H\"{o}gele, A.}, \bibinfo{author}{F\"{a}lt, S.}
   \& \bibinfo{author}{Imamo\u{g}lu, A.}
\newblock \bibinfo{title}{Observation of Dressed Ecitonic States in a Single Quantum Dot}.
\newblock \emph{\bibinfo{journal}{Phys. Rev. Lett.}}
  \textbf{\bibinfo{volume}{100}}, \bibinfo{pages}{177401}
  (\bibinfo{year}{2008}).

\bibitem{Muller.Flagg:2007}
\bibinfo{author}{Muller, A.}, \emph{et~al.}
\newblock \bibinfo{title}{Resonance Fluorescence from a Coherently Driven Semiconductor Quantum Dot in a Cavity}.
\newblock \emph{\bibinfo{journal}{Phys. Rev. Lett.}}
  \textbf{\bibinfo{volume}{99}}, \bibinfo{pages}{187402}
  (\bibinfo{year}{2007}).

\bibitem{Vamivakas.Atature:2009}
\bibinfo{author}{Vamivakas, A. N.}, \bibinfo{author}{Zhong, Y.},
  \bibinfo{author}{Yong, C.-Y.} \& \bibinfo{author}{Atature, M.}
\newblock \bibinfo{title}{Spin-resolved quantum-dot resonance fluorescence}.
\newblock \emph{\bibinfo{journal}{Nature Physics}}
  \textbf{\bibinfo{volume}{5}}, \bibinfo{pages}{198-202}
  (\bibinfo{year}{2009}).

\bibitem{Muller.Fang:2009}
\bibinfo{author}{Muller, A.}, \bibinfo{author}{Fang, W.},
  \bibinfo{author}{Lawall, J.} \& \bibinfo{author}{Solomon, G. S.}
\newblock \bibinfo{title}{Creating Polarization-Entangled Photon Pairs from a Semiconductor Quantum Dot Using the Optical Stark Effect}.
\newblock \emph{\bibinfo{journal}{Phys. Rev. Lett.}}
  \textbf{\bibinfo{volume}{103}}, \bibinfo{pages}{217402}
  (\bibinfo{year}{2009}).

\bibitem{Nienhuis:1993}
\bibinfo{author}{Nienhuis, G.}
\newblock \bibinfo{title}{Spectral correlations in resonance fluorescence}.
\newblock \emph{\bibinfo{journal}{Phys. Rev. A}}
  \textbf{\bibinfo{volume}{47}}, \bibinfo{pages}{510-518}
  (\bibinfo{year}{1993}).

\bibitem{Tannoudji.Grynberg:2004}
\bibinfo{author}{Cohen-Tannoudji, C.}, \bibinfo{author}{Dupont-Roc, J.} \&
  \bibinfo{author}{Grynberg, G.}
\newblock \emph{\bibinfo{title}{Atom-Photon Interactions}}
  (\bibinfo{publisher}{Wiley-VCH Weinheim}, \bibinfo{year}{2004}).

\bibitem{Aichele.Benson:2004}
\bibinfo{author}{Aichele, T.}, \bibinfo{author}{Reinaudi, G.}
 \& \bibinfo{author}{Benson, O.}
\newblock \bibinfo{title}{Seperating cascaded photons from a single quantum dot: Demonstration of multiplexed quantum cryptography}.
\newblock \emph{\bibinfo{journal}{Phys. Rev. B}}
  \textbf{\bibinfo{volume}{70}}, \bibinfo{pages}{235329}
  (\bibinfo{year}{2004}).

\bibitem{Santori.Pelton:2001}
\bibinfo{author}{Santori, C.}, \bibinfo{author}{Pelton, M.},
\bibinfo{author}{Salamo, G.}, \bibinfo{author}{Dale, Y.}
 \& \bibinfo{author}{Yamamoto, Y.}
\newblock \bibinfo{title}{Triggered Single Photons from a Quantum Dot}.
\newblock \emph{\bibinfo{journal}{Phys. Rev. Lett.}}
  \textbf{\bibinfo{volume}{86}}, \bibinfo{pages}{1502}
  (\bibinfo{year}{2001}).

\bibitem{Bayer:2002}
\bibinfo{author}{Bayer, M.}, \emph{et~al.}
\newblock \bibinfo{title}{Fine structure of neutral and charged excitons in self-ssembled In(Ga)As/(Al)GaAs quantum dots}.
\newblock \emph{\bibinfo{journal}{Phys. Rev. B}}
  \textbf{\bibinfo{volume}{65}}, \bibinfo{pages}{195315}
  (\bibinfo{year}{2002}).

\bibitem{Akopian.Rastelli:2011}
\bibinfo{author}{Akopian, N.}, \bibinfo{author}{Wang, L.},
  \bibinfo{author}{Rastelli, A.}, \bibinfo{author}{Schmidt O. G}
   \& \bibinfo{author}{Zwiller, V.}
\newblock \bibinfo{title}{Hybrid semiconductor-atomic interface: slowing down single photons from a quantum dot}.
\newblock \emph{\bibinfo{journal}{Nat. Phot.}}
  \textbf{\bibinfo{volume}{5}}, \bibinfo{pages}{230-233}
  (\bibinfo{year}{2011}).

\bibitem{Gisin.Thew:2007}
\bibinfo{author}{Gisin, N.} \& \bibinfo{author}{Thew, R.}
\newblock \bibinfo{title}{Quantum Communication}.
\newblock \emph{\bibinfo{journal}{Nat. Phot.}}
  \textbf{\bibinfo{volume}{1}}, \bibinfo{pages}{165-171}
  (\bibinfo{year}{2007}).

\bibitem{Freedhoff.Quang:1994}
\bibinfo{author}{Freedhoff, H.} \& \bibinfo{author}{Quang, T.}
\newblock \bibinfo{title}{Ultrasharp Lines in the Absorption and Fluorescence Spectra of an Atom in a Cavity}.
\newblock \emph{\bibinfo{journal}{Phys. Rev. Lett.}}
  \textbf{\bibinfo{volume}{72}}, \bibinfo{pages}{474}
  (\bibinfo{year}{1994}).

\bibitem{Quang. Freedhoff:1993}
\bibinfo{author}{Quang, T.} \& \bibinfo{author}{Freedhoff, H.}
\newblock \bibinfo{title}{Atomic population inversion and enhancement of resonance fluorescence in a cavity}.
\newblock \emph{\bibinfo{journal}{Phys. Rev. A.}}
  \textbf{\bibinfo{volume}{47}}, \bibinfo{pages}{2285}
  (\bibinfo{year}{1993}).

\end{thebibliography}

\begin{thebibliography}{10}
\expandafter\ifx\csname url\endcsname\relax
  \def\url#1{\texttt{#1}}\fi
\expandafter\ifx\csname
urlprefix\endcsname\relax\def\urlprefix{URL }\fi
\providecommand{\bibinfo}[2]{#2}
\providecommand{\eprint}[2][]{\url{#2}}

\bibitem{Aichele:2004}
\bibinfo{author}{Aichele, T.}
\newblock \emph{\bibinfo{title}{Detection and Generation of Non-Classical Light States from Single Quantum Emitters (PhD thesis)}}
  (\bibinfo{publisher}{Logos}, \bibinfo{year}{2005}).

\bibitem{Nienhuis.Nienhuis:1993}
\bibinfo{author}{Nienhuis, G.}
\newblock \emph{\bibinfo{journal}{Phys. Rev. A}}
  \textbf{\bibinfo{volume}{47}}, \bibinfo{pages}{510-518}
  (\bibinfo{year}{1993}).

\bibitem{Santori.Salamo:2001}
\bibinfo{author}{Santori, C.}, \emph{et al.}
\newblock \emph{\bibinfo{journal}{Phys. Rev. Lett.}}
  \textbf{\bibinfo{volume}{86}}, \bibinfo{pages}{1502}
  (\bibinfo{year}{2001}).

\end{thebibliography}
\end{document}